\documentclass[aip,preprint]{revtex4-1}   

\usepackage{graphicx}
\usepackage{amsmath}  

\begin{document}

\title[Self-excitation of PSR oscillations in CCRF discharges]{On the self-excitation mechanisms of Plasma Series Resonance oscillations in single- and multi-frequency capacitive discharges}


\author{Edmund Sch\"ungel}
\affiliation{Department of Physics, West Virginia University, Morgantown, West Virginia 26506-6315, USA} 

\author{Steven Brandt}
\affiliation{Department of Physics, West Virginia University, Morgantown, West Virginia 26506-6315, USA} 

\author{Ihor Korolov}
\affiliation{Institute for Solid State Physics and Optics, Wigner Research Centre for Physics, Hungarian Academy of Sciences, 1121 Budapest, 
Konkoly-Thege Mikl\'os str. 29-33, Hungary}

\author{Aranka Derzsi}
\affiliation{Institute for Solid State Physics and Optics, Wigner Research Centre for Physics, Hungarian Academy of Sciences, 1121 Budapest, 
Konkoly-Thege Mikl\'os str. 29-33, Hungary}

\author{Zolt\'an Donk\'o}
\affiliation{Institute for Solid State Physics and Optics, Wigner Research Centre for Physics, Hungarian Academy of Sciences, 1121 Budapest, 
Konkoly-Thege Mikl\'os str. 29-33, Hungary}

\author{Julian Schulze}
\affiliation{Department of Physics, West Virginia University, Morgantown, West Virginia 26506-6315, USA} 

\begin{abstract}
The self-excitation of plasma series resonance (PSR) oscillations is a prominent feature in the current of low pressure capacitive radio frequency (RF) discharges. This resonance leads to high frequency oscillations of the charge in the sheaths and enhances electron heating. Up to now, the phenomenon has only been observed in asymmetric discharges. There, the nonlinearity in the voltage balance, which is neccessary for the self-excitation of resonance oscillations with frequencies above the applied frequencies, is caused predominantly by the quadratic contribution to the charge-voltage relation of the plasma sheaths. Using PIC/MCC simulations of single- and multi-frequency capacitive discharges and an equivalent circuit model, we demonstrate that other mechanisms such as a cubic contribution to the charge-voltage relation of the plasma sheaths and the time dependent bulk electron plasma frequency can cause the self-excitation of PSR oscillations, as well. These mechanisms have been neglected in previous models, but are important for the theoretical description of the current in symmetric or weakly asymmetric discharges.
\end{abstract}


\maketitle 

\section{Introduction}

The nonlinear nature of capacitively coupled radio frequency (CCRF) plasmas leads to a complex physical behavior. In particular, the generation of higher harmonics in the discharge current of a CCRF plasma driven by a sinusoidal voltage (e.g. at 13.56 MHz) has been a subject of scientific interest, as the shape of the current waveform has a direct impact on the electron heating dynamics \cite{PSR_Lieberman2,PSR_stoch_Schulze,PSR_Ziegler2} and, therefore, on the plasma sustainment via ionization and on the densities, fluxes, and eventually process rates in surface processing applications \cite{PSR_Yamazawa,PSR_Klick}. The plasma series resonance (PSR) circuit is formed by the combination of the electron inertia in the plasma bulk and the nonlinear capacitances of the sheaths between the plasma and the surrounding surfaces. Due to this nonlinearity present in the series circuit \cite{sheath_nonlinearity_Klick,PSR_nonlinear_Mussenbrock}, the PSR can be self-excited in low pressure CCRF discharges, although the typical PSR frequency is well above the frequency of the driving voltage \cite{PSR_model_UCZ}. 
Such PSR oscillations have not yet been studied intensively in geometrically symmetric discharges, but have been studied extensively in many experiments, simulations, and theoretical models of asymmetric discharges (see, e.g. Refs. \onlinecite{PSR_Allen1,PSR_Klick,PSR_Birdsall1,PSR_model_UCZ,PSR_Mussenbrock,PSR_nonlinear_Mussenbrock,PSR_Semmler,PSR_diagnostics_Schulze,PSR_Ziegler1,
PSR_stoch_Schulze,PSR_Lieberman2,PSR_EAE_Zoltan,PSR_Yamazawa,PSR_Ziegler2,PSR_Wang,PSR_Bora2,PSR_Bora3,Bora4,Bora1,PSR_EAE_geomasymm,sheath_nonlinearity_Klick} and references therein), where the asymmetry is typically caused by the chamber geometry: the unequal surface areas of the powered and grounded electrodes.

Another way of inducing an asymmetry is via the Electrical Asymmetry Effect (EAE) \cite{PSR_EAE_escampig,EAE2,PSR_EAE_PIC,PSR_EAE_Zoltan,PSR_EAE_Qdyn,PSR_Bora3,Bora4,Bora1,PSR_EAE_geomasymm,EAEbienholz}, i.e. by using a driving voltage waveform with unequal absolute values of the global maximum and minimum. Donk\'o \textit{et al.}\cite{PSR_EAE_Zoltan} have shown that the PSR is self-excited in geometrically symmetric, electrically asymmetric discharges operated at a fundamental frequency (13.56 MHz) and its second harmonic (27.12 MHz). The control of the discharge asymmetry via the EAE, i.e. by tuning the relative phase between the applied harmonics, translates into a convenient control of the DC self-bias, mean sheath voltages, and mean ion energy at the electrodes independently of the ion flux. The range of this control can be extended by using multiple consecutive harmonics \cite{EAEmultif1,EAEmultif2,EAEjohnson,EAEjohnson3}. Moreover, an asymmetry can be caused by applying a temporally asymmetric sawtooth waveform due to the difference in the upward and downward slopes \cite{Lafleur_asymm1,Lafleur_asymm2}.

An asymmetry was believed to be a neccessary requirement for the self-excitation of PSR oscillations \cite{PSR_Bora2}, because the nonlinearity of the sheaths will cancel in a symmetric discharge if they are modelled by a quadratic charge-voltage relation \cite{sheath_nonlinearity_Klick,PSR_nonlinear_Mussenbrock,PSR_model_UCZ}. 
Here, we demonstrate that other mechanisms can cause the self-excitation of the PSR even in symmetric or weakly asymmetric CCRF plasmas based on simulations of discharges excited by a single-frequency voltage waveform, $\phi_{\sim}(\varphi)=\phi_{tot} \, \textnormal{cos}(k \varphi)$ with $k=1$ or $k=4$, or by a superposition of four harmonics \cite{EAEmultif1,EAEmultif2},
\begin{equation}
\label{AppVol}
\phi_{\sim}(\varphi)=\phi_{tot}  \sum_{k=1}^4 \frac{5-k}{10} \textnormal{cos}(k \varphi + \theta_k),
\end{equation}
where $\phi_{tot}$ is the applied voltage amplitude and $\varphi=\omega t$ ($\omega = 2 \pi$ 13.56 $\times$ $10^6$ rad s$^{-1}$) is the RF phase (see figure \ref{fig_voltages}). The relative amplitudes in equation (\ref{AppVol}) are optimized to provide the greatest control range of the DC self-bias as a function of the phase shifts \cite{EAEmultif1}. In this work, we use $\theta_1=\theta_3=0$ and $\theta_2=\theta_4=\pi/2$, so that the absolute values of the maximum and the minimum of $\phi_{\sim}(\varphi)$ are equal. Hence, the resulting profiles of the density and potential averaged over the RF period are approximately symmetric around the discharge center; there is only a slight disturbance in the symmetry of the discharge due to the weak asymmetry of the applied multi-frequency voltage waveform in time \cite{Lafleur_asymm1,Lafleur_asymm2}.
Using an equivalent circuit model, the self-excitation will be shown to be facilitated by the nonlinearity, which can be attributed to the cubic contribution to the charge-voltage relation of the plasma sheaths, and to the time dependent bulk electron plasma frequency.

\section{Description of simulation and model}

\begin{figure}[tbp]
\includegraphics[width=0.4775\textwidth]{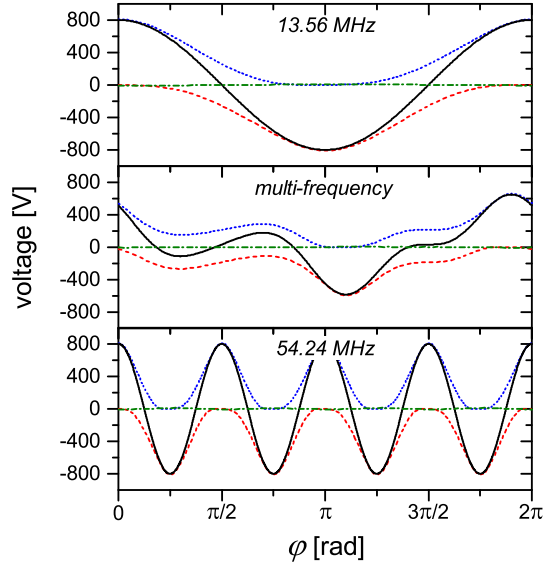}
\caption{Total voltage drop between the electrodes, i.e. applied voltage and DC self-bias (black solid line), and individual voltage drops across the powered electrode sheath (red dashed line), the grounded electrode sheath (blue dotted line), and the plasma bulk (green dash-dotted line), respectively. The discharge is driven by a 13.56 MHz (top), a multifrequency (equation (\ref{AppVol}), middle), and a 54.24 MHz (bottom) voltage waveform, respectively.} 
\label{fig_voltages}
\end{figure}

The 1d3v particle in cell simulation code, which is complemented with a Monte Carlo treatment of collision processes (PIC/MCC) describes geometrically symmetric discharges, as the electrodes are assumed to be infinite, planar, and parallel. 
One of the electrodes is driven by the prescribed voltage waveform (see above), while the other electrode located at a distance of $d=$30 mm is grounded. 
The discharge is operated at $\phi_{tot}=800$ V in argon. The cross section data set includes elastic, excitation, and ionization collisions of electrons with Ar atoms, as well as isotropic and backward elastic scattering of argon ions from Ar atoms \cite{PhelpsOnline,Phelps1,Phelps2}. The neutral gas pressure and temperature are 3 Pa and 400 K, respectively. Both the ion induced secondary electron emission coefficient and the electron reflection coefficient are taken to be 0.2 at both electrodes. 

in the simulations, the discharge is driven by a voltage source, i.e. the potential at the powered electrode is specified as the sum of the applied RF voltage and the DC self-bias. No assumption on the outer driving circuit is necessary. Hence, the simulation is capable of reproducing ideal experimental configurations of CCRF plasmas, where the voltage at the powered electrode is measured and any influence of the external driving circuit on the plasma is irrelevant.

\begin{figure}[tbp]
\includegraphics[width=0.4775\textwidth]{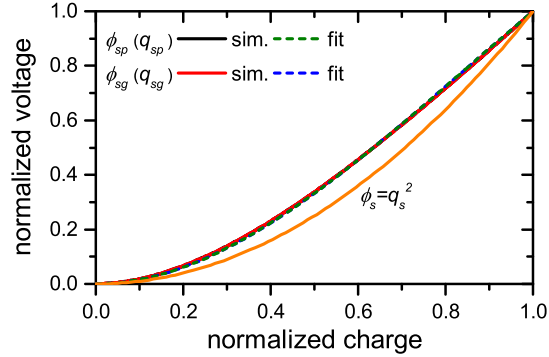}
\caption{Voltage drop across the powered and grounded electrode sheath, $\phi_{sp}$ and $\phi_{sg}$, as a function of the charge within the respective sheath resulting from the PIC/MCC simulation (sim.) at 54.24 MHz and model functions (\ref{Usp}) and (\ref{Usg}), fitted to the simulation data (fit). Here, $a \approx$1.66 and $b \approx$1.66, respectively. A simple quadratic relation is shown, as well.}
\label{fig_Q_V_relation}
\end{figure}

The model, which is used to gain an understanding of the underlying physical mechanisms, is based on a theory developed by Czarnetzki \textit{et al} \cite{PSR_model_UCZ}. The sum of the applied voltage, $\phi_{\sim}(\varphi)$, and the DC self-bias, $\eta$, drop across the CCRF discharge and, therefore, have to balance the sum of the individual voltage drops across the powered and grounded electrode sheath, $\phi_{sp}$ and $\phi_{sg}$, and across the plasma bulk, $\phi_{bulk}$: \cite{PSR_model_UCZ,PSR_EAE_escampig}
\begin{equation}
\label{VolBal}
\bar{\eta} + \bar{\phi}_{\sim}(\varphi)=\phi_{sp} (\varphi) + \phi_{bulk}(\varphi) + \phi_{sg}(\varphi).
\end{equation}
Here, all quantities are normalized by the applied voltage amplitude, $\phi_{tot}$. 
Now, the voltages over the sheaths and the bulk are described by their functional dependence on the charge in the powered electrode sheath, $Q(\varphi)$. For the two sheaths, a cubic charge-voltage relation is adapted from Ref. \onlinecite{UCZ_sheath_model}:
\begin{eqnarray}
\label{Usp}
\phi_{sp}(\varphi) & = & -q_{tot}^2 q^2(\varphi) \left[ q(\varphi) (1-a) +a \right],  \\
\phi_{sg}(\varphi) & = & \varepsilon q_{tot}^2 \left[ 1- q(\varphi) \right]^2 \left[ [1-q(\varphi)] (1-b) +b \right].
\label{Usg}
\end{eqnarray}
The total uncompensated charge, $q_{tot}$, is constant and defined in a way that the oscillating charge varies between $q=1$ (fully expanded sheath) and $q=0$ (fully collapsed sheath, i.e. neglecting the floating potential). Accordingly, $q(\varphi)=Q(\varphi)/Q_0$ with $Q_0=e \int_0^{s_{max}} n_i(z) dz$, where $n_i(z)$ is the ion density profile and $s_{max}$ is the maximum powered sheath length. The symmetry parameter $\varepsilon$ is defined as \cite{EAE2,PSR_EAE_escampig}
\begin{equation}
\label{SymPar}
\varepsilon = \left| \frac{\phi^{max}_{sg}}{\phi^{max}_{sp}} \right|.
\end{equation}
In contrast to all previous investigations on the self-excited PSR, where $\varepsilon \lesssim 0.5$ and the asymmetry has been assumed to be a neccessary condition for the self-excitation, the symmetry parameter is unity or close to unity in all cases considered here ($\varepsilon$=1.00, 1.00, and 1.11 for the 13.56 MHz, 54.24 MHz, and multi-frequency voltage waveform, respectively), as we only investigate symmetric or weakly asymmetric discharges in this work. $a$ and $b$ are cubic correction parameters; $a=1$ and $b=1$ yields simple quadratic charge-voltage relations, so that the sum $\phi_{sp}+\phi_{sg}$ is a linear function of $q$ in a symmetric discharge, i.e. when $\varepsilon$ equals unity and all non-linearities cancel. However, the cubic term caused by the inhomogeneity in the static ion sheath density profile \cite{UCZ_sheath_model} does not cancel out in symmetric discharges and must not be neglected. As the sheath charge, $Q$, is determined by the ion density profile, $Q(\varphi)=e \int_0^{s(\varphi)} n_i(z) dz$ where $s(\varphi)$ is the plasma sheath edge, and the sheath voltage is proportional to $\int_0^{s(\varphi)} Q(\varphi) dz$ via Poisson's equation \cite{EAE2}, the charge-voltage relation varies as a function of the RF phase and, hence, leads to a deviation from a simple quadratic relationship. Taking this severe deviation (see figure \ref{fig_Q_V_relation}) into account is important for a correct treatment of the sheath voltage. Thus, $a \neq 1$ and $b \neq 1$. 

These parameters are obtained from fits of $\phi_{sp}/q_{tot}^2$ as a function of $q$ and $\phi_{sg}/(\varepsilon q_{tot}^2)$ as a function of $q_g=1-q$ (see equations (\ref{Usp}) and (\ref{Usg})) to the normalized charge-voltage relation resulting from the simulations with excellent agreement, as it is shown in figure \ref{fig_Q_V_relation}. Typically, we find $a \approx 1.5 - 1.7$ and $b \approx a$.
The charge-voltage relations of the sheaths are obtained from the simulations by (i) taking the difference of the potential between the momentary plasma sheath edges, which are determined according to a criterion defined in Ref. \onlinecite{Brinkmann}, and the electrode surfaces and by (ii) adding up all net charges in these momentary sheath regions.


\begin{figure}[tbp]
\includegraphics[width=0.4775\textwidth]{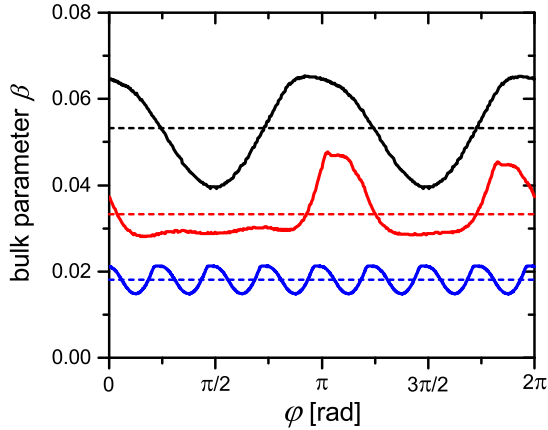}
\caption{Bulk parameter, $\beta$, as a function of the RF phase for the 13.56 MHz (black), multi-frequency (equation (\ref{AppVol}), red), and 54.24 MHz (blue) discharge. The dashed lines correspond to RF period averaged values.}
\label{fig_beta}
\end{figure}

The voltage drop across the plasma bulk derived from the electron momentum balance equation (see Ref. \onlinecite{PSR_model_UCZ} for details) consists of two terms,
\begin{equation}
\label{BulkVoltage}
\phi_{bulk}(\varphi) =- 2 \beta^2(\varphi) \left[ \ddot{q}(\varphi) + \kappa \dot{q} (\varphi) \right],
\end{equation}
representing an inductance and a resistance due to electron inertia and electron momentum transfer collisions, respectively. Here, a dot denotes differentiation with respect to $\varphi$, and $\kappa=\nu_m/\omega$ is the normalized electron collision frequency. In agreement with the recent findings of \textit{Lafleur et al.} for discharges operated at similar conditions \cite{Lafleur_collision}, we use a value of $\nu_m \approx 2.0 \times 10^8$ s$^{-1}$ or $\kappa \approx 2.4$ in the model. The bulk parameter $\beta$ is defined as \cite{PSR_model_UCZ}
\begin{equation}
\beta(\varphi)=\frac{\omega}{\omega_{pe}(\varphi)}  \sqrt{ \frac{L_{bulk}(\varphi)}{s_{max}}}
\end{equation}
and depends on the ratio of the bulk length, $L_{bulk}$, to the maximum sheath extension, $s_{max}$, and the ratio of the angular frequency of the applied (fundamental) frequency, $\omega$, to an effective electron plasma frequency in the plasma bulk region, which is defined as
\begin{equation}
\omega_{pe}^{-1}(\varphi)=\sqrt{ \frac{\epsilon_0 m_e}{e^2 L_{bulk}(\varphi)} \int_{s_p(\varphi)}^{d-s_g(\varphi)} n_e^{-1}(\varphi,z) dz},
\end{equation}
where $\epsilon_0$, $e$, $m_e$, and $n_e$ are the vacuum permittivity and the charge, the mass, and the density of the electrons. Note that - in contrast to previous works on the PSR in strongly asymmetric CCRF discharges \cite{PSR_model_UCZ} - we allow for a time dependence of the bulk parameter $\beta(\varphi)$, i.e. it is calculated by integrating the inverse electron density profile along the bulk length $L_{bulk}(\varphi)$ between the momentary positions of the plasma sheath edges. The temporal evolution of $\beta(\varphi)$ within the RF period is shown in figure \ref{fig_beta}. It is evident that it changes significantly in all three cases. This modulation is due to the fact that the electron density follows the ion density profile within the bulk, whereas the sheath to bulk transitions sweep over regions of different densities due to the inhomogeneous ion density profiles in the sheaths. Therefore, the effective electron plasma frequency decreases and, as a result, $\beta$ increases strongly during either sheath collapse. 
Apparently, the change is much stronger in the multi-frequency case due to the relatively short times of collapsing sheaths \cite{PSR_EAE_Qdyn} (see figure \ref{fig_voltages}). Finally, equations (\ref{Usp}), (\ref{Usg}), and (\ref{BulkVoltage}) are inserted into equation (\ref{VolBal}) and the resulting equation is solved numerically for the charge $q$. The electron current is then calculated in the model by taking the first derivative of $q$ with respect to the RF phase and multiplying it with the factor $\omega Q_0$ (determined from the simulations) for denormalization.

\begin{figure*}[htbp]
\includegraphics[width=0.96\textwidth]{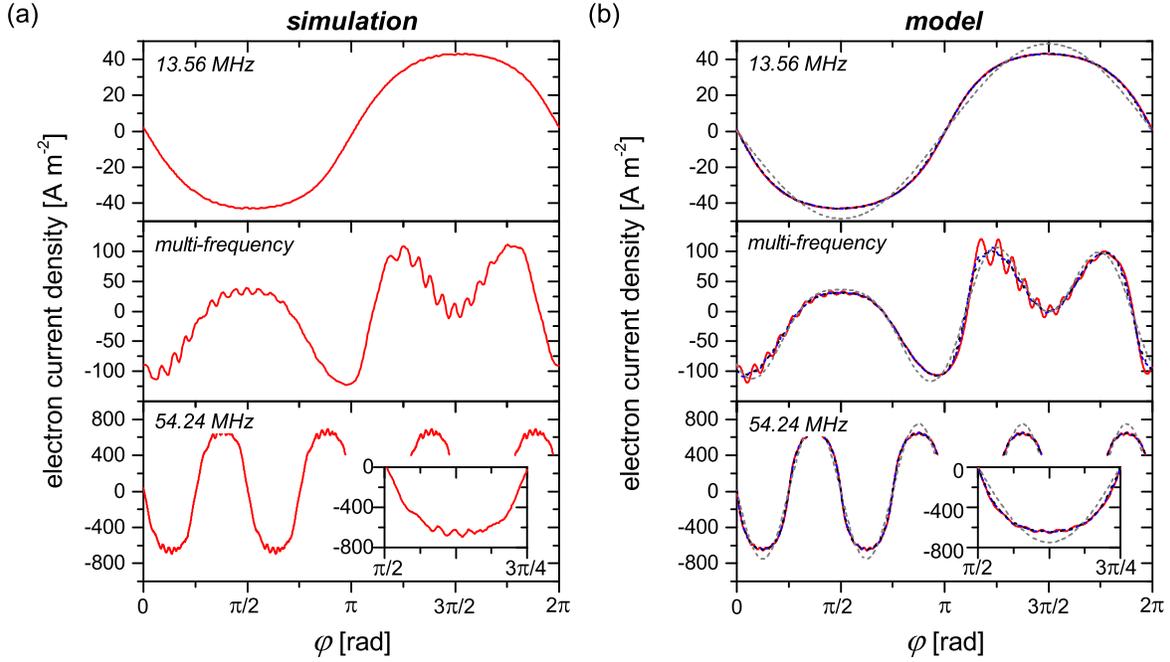}
\caption{Electron current in the plasma bulk obtained from (a) the PIC/MCC simulations and (b) the PSR model. The discharge is driven by a 13.56 MHz (top), a multi-frequency (equation (\ref{AppVol}), middle), and a 54.24 MHz (bottom) voltage waveform, respectively. The model plots show the current obtained using $\beta(\varphi)$, $a$, and $b$ (red solid line), $\beta(\varphi)$ with $a=1$ and $b=1$ (dashed gray line), or the RF period averaged $\beta$, $a$, and $b$ (blue dashed line) from the PIC/MCC simulations. The black dotted lines correspond to solutions neglecting the bulk voltage.}
\label{fig_current}
\end{figure*}

\section{Results}

Figure \ref{fig_current} shows the electron current flowing through the discharge center obtained from the PIC/MCC simulations and the PSR model for three different cases. The plasma series resonance is not self-excited in a single-frequency discharge driven by a 13.56 MHz voltage.
By comparing the current resulting from the model using the time-resolved $\beta(\varphi)$ with the model current without PSR, i.e. neglecting the bulk voltage, one finds that the perturbation due to the PSR is less than 1 \%. Accordingly, all model curves are almost identical. 

In the multi-frequency case, i.e. when the discharge is operated at a voltage waveform according to equation (\ref{AppVol}), the excitation of the PSR is relatively strong. The result of the model matches the simulated current if and only if the temporal variation of $\beta(\varphi)$ obtained from the simulations is taken into account. This might explain the discrepancy between the theoretical and the measured current waveforms observed in previous studies \cite{PSR_Klick, PSR_Ziegler1,PSR_EAE_escampig}. Different from these investigations of geometrically asymmetric discharges, figure \ref{fig_current} shows that the PSR is self-excited at the times of the collapse of both the powered and the grounded electrode sheath. 

A single-frequency discharge driven by a 54.24 MHz voltage exhibits a plasma series resonance; however, the excitation and amplitude of the higher harmonics are much smaller compared to the multi-frequency discharge. This is mainly due to the fact that both the absolute value and the variation of the bulk parameter $\beta(\varphi)$ are smaller (see figure \ref{fig_beta}), because the plasma density and, hence, the effective electron plasma frequency are higher and the RF period accumulated time of sheath collapse is longer. Note that the total current amplitude is also much larger. Furthermore, only odd harmonics of the applied frequency are allowed in the current of a symmetric discharge \cite{sheath_nonlinearity_Klick}. Therefore, the application of a multi-frequency voltage waveform generally allows for a much broader Fourier spectrum in the discharge current. Nevertheless, the inlay zooming into a minimum of the current within the RF period depicted in figure \ref{fig_current} reveals a weak resonance.  Moreover, the model curves show that the overall shape of the current waveform will be different and that no high frequency oscillations will be observed, if a simple quadratic sheath charge-voltage relation is used. Accordingly, we find that both the temporal dependence of $\beta(\varphi)$ and the deviation from a quadratic sheath charge-voltage relation are crucial for the self-excitation of the PSR and must be carefully considered in order to correctly describe the effect. This is particularly important as the shape of the current waveform is essential for the description of the electron heating and power dissipation in CCRF plasmas \cite{PSR_Lieberman2,PSR_stoch_Schulze,PSR_Ziegler2}.

\begin{figure}[tp]
\includegraphics[width=0.49\textwidth]{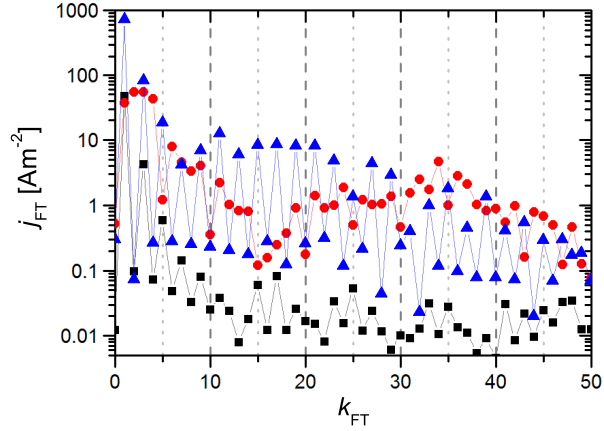}
\caption{Fourier amplitude spectrum, $j_{\textnormal{FT}}$, of the electron current density obtained from PIC/MCC simulations of discharges operated at 13.56 MHz (black squares), a multi-frequency voltage (equation (\ref{AppVol}), red dots), and 54.24 MHz (blue triangles).}
\label{fig_FFT}
\end{figure}

Figure \ref{fig_FFT} shows the Fourier amplitude spectrum of the electron current density obtained from PIC/MCC simulations. Here, $k_{\textnormal{FT}}$ is the harmonic number with respect to the lowest applied frequency, i.e. $k_{\textnormal{FT}}=1$ corresponds to the 13.56 MHz component in the analysis of the discharges driven by 13.56 MHz and by the multi-frequency waveform and it corresponds to the 54.24 MHz component in the analysis of the discharges driven by 54.24 MHz. Note the logarithmic scale on the vertical axis. In the discharge driven by 13.56 MHz, the contribution of higher harmonics drops quickly as a function of $k_{\textnormal{FT}}$. The odd harmonics are generally more pronounced compared to the even harmonics due to the symmetry constraints of the current waveform \cite{sheath_nonlinearity_Klick}. In electrically asymmetric discharges driven by multiple consecutive harmonics, such constraints are no longer given and a broad spectrum of higher harmonics is observed due to the self-excitation of the PSR \cite{Bora1,PSR_Bora2,PSR_Bora3,PSR_Lieberman2,PSR_Wang}. The Fourier spectrum of the electron current density of the 54.24 MHz discharge again reveals the alternating amplitudes for odd and even harmonics typical for symmetric discharges, but the contribution of the odd harmonics remains significant within a large range of $k_{\textnormal{FT}}$. In particular, the PSR oscillations observed in the bottom panel of figure \ref{fig_current}(a) mainly correspond to the plateau of the odd harmonics between $k_{\textnormal{FT}}=13$ and $k_{\textnormal{FT}}=23$.

\section{Conclusions}

In conclusion, the self-excitation of plasma series resonance oscillations is investigated in single- and multi-frequency capacitive discharges using a combined approach of self-consistent PIC/MCC simulations and an appropriate model. The mechanisms behind the PSR self-excitation in symmetric or weakly asymmetric plasmas, i.e. the cause of the required nonlinearity, are found to be a change in the bulk parameter $\beta(\varphi)$ as a function of the RF phase and a cubic component in the charge-voltage relation of the sheaths. The former is caused by a sweep of the plasma bulk region across a strongly varying electron density profile, i.e. the electron density follows the decrease of the ion density profile from the presheath region to the temporal sheath edge. Therefore, the inductance in the series circuit becomes time dependent. Regarding the latter, the sum of the sheath charge-voltage relations remains nonlinear even in a symmetric discharge due to a cubic term, which is related to the inhomogeneous ion density profile in the sheath region. These mechanisms of PSR self-excitation must not be neglected in symmetric and weakly asymmetric CCRF plasmas, as the model will reproduce the simulated current only if these effects are implemented correctly, e.g. by taking the respecitve parameters obtained from simulations. The effect of the self-excitation of the PSR on the electron heating dynamics in symmetric CCRF plasmas will be addressed in a future study.

\section*{Acknowledgements}

We thank James Franek (West Virginia University) and Uwe Czarnetzki (Ruhr-University Bochum) for helpful discussions. Funding by the Hungarian Scientific Research Fund through the grant OTKA K-105476 is gratefully acknowledged.

\end{document}